\documentclass[8.5pt,twoside,twocolumn]{article}
\usepackage[super,sort&compress,comma]{natbib} 
\usepackage{mhchem}
\usepackage{times,mathptmx}
\usepackage{sectsty}
\usepackage{balance} 

\usepackage{graphicx} %eps figures can be used instead
\usepackage{lastpage}
\usepackage[format=plain,justification=raggedright,singlelinecheck=false,font=small,labelfont=bf,labelsep=space]{caption} 
\usepackage{fancyhdr}
\begin{document}

\thispagestyle{plain}
\fancypagestyle{plain}{
\fancyhead[L]{\includegraphics[height=8pt]{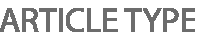}}
\fancyhead[C]{\hspace{-1cm}\includegraphics[height=20pt]{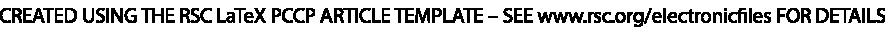}}
\fancyhead[R]{\includegraphics[height=10pt]{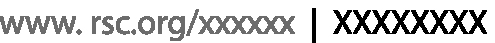}\vspace{-0.2cm}}
\renewcommand{\headrulewidth}{1pt}}
\renewcommand{\thefootnote}{\fnsymbol{footnote}}
\renewcommand\footnoterule{\vspace*{1pt}% 
\hrule width 3.4in height 0.4pt \vspace*{5pt}} 
\setcounter{secnumdepth}{5}

\makeatletter 
\def\subsubsection{\@startsection{subsubsection}{3}{10pt}{-1.25ex plus -1ex minus -.1ex}{0ex plus 0ex}{\normalsize\bf}} 
\def\paragraph{\@startsection{paragraph}{4}{10pt}{-1.25ex plus -1ex minus -.1ex}{0ex plus 0ex}{\normalsize\textit}} 
\renewcommand\@biblabel[1]{#1}            
\renewcommand\@makefntext[1]% 
{\noindent\makebox[0pt][r]{\@thefnmark\,}#1}
\makeatother 
\renewcommand{\figurename}{\small{Fig.}~}
\sectionfont{\large}
\subsectionfont{\normalsize} 

\fancyfoot{}
\fancyfoot[LO,RE]{\vspace{-7pt}\includegraphics[height=9pt]{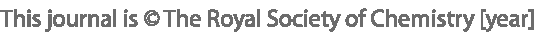}}
\fancyfoot[CO]{\vspace{-7.2pt}\hspace{12.2cm}\includegraphics{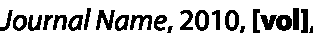}}
\fancyfoot[CE]{\vspace{-7.5pt}\hspace{-13.5cm}\includegraphics{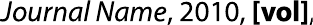}}
\fancyfoot[RO]{\footnotesize{\sffamily{1--\pageref{LastPage} ~\textbar  \hspace{2pt}\thepage}}}
\fancyfoot[LE]{\footnotesize{\sffamily{\thepage~\textbar\hspace{3.45cm} 1--\pageref{LastPage}}}}
\fancyhead{}
\renewcommand{\headrulewidth}{1pt} 
\renewcommand{\footrulewidth}{1pt}
\setlength{\arrayrulewidth}{1pt}
\setlength{\columnsep}{6.5mm}
\setlength\bibsep{1pt}

\twocolumn[
  \begin{@twocolumnfalse}
\noindent\LARGE{\textbf{Avalanches, Plasticity, and Ordering in Colloidal 
Crystals Under Compression }}
\vspace{0.6cm}

\noindent\large{\textbf{Danielle McDermott,\textit{$^{1,2}$}  
Cynthia J. Olson Reichhardt,$^{\ast}$\textit{$^{1}$} 
and Charles Reichhardt\textit{$^{1}$}}}\vspace{0.5cm}

\noindent\textit{\small{\textbf{Received Xth XXXXXXXXXX 20XX, Accepted Xth XXXXXXXXX 20XX\newline
First published on the web Xth XXXXXXXXXX 200X}}}

\noindent \textbf{\small{DOI: 10.1039/b000000x}}
\vspace{0.6cm}
%Please do not change this text.

\noindent \normalsize{
Using numerical simulations we examine 
colloids with a long-range Coulomb interaction
confined in a two-dimensional trough potential undergoing 
dynamical compression.  
As the depth of the confining well is increased, 
the colloids move via
elastic distortions interspersed with  
intermittent bursts or avalanches of plastic motion.
In these avalanches, the colloids rearrange
to minimize their colloid-colloid repulsive interaction 
energy by adopting an
average lattice constant 
that is isotropic 
despite the anisotropic nature of the compression. 
The avalanches   
take the form of shear banding events 
that decrease or increase the structural order of the system. 
At larger compressions, the avalanches are associated 
with a reduction of the number of rows of colloids that fit within the 
confining potential, and between avalanches
the colloids can exhibit partially crystalline or 
even smectic ordering. 
The colloid velocity distributions during the avalanches 
have a non-Gaussian form with power law tails and exponents that
are consistent with those found for 
the velocity distributions of gliding dislocations. 
We observe similar
behavior when we subsequently decompress the system, 
and find a partially hysteretic response reflecting the irreversibility of the 
plastic events.    
}
\vspace{0.5cm}
 \end{@twocolumnfalse}
  ]

\section{Introduction}

\footnotetext{\textit{$^{1}$~Theoretical Division,
Los Alamos National Laboratory, Los Alamos, New Mexico 87545, USA.
Fax: 1 505 606 0917; Tel: 1 505 665 1134; E-mail: cjrx@lanl.gov}}
\footnotetext{\textit{$^{2}$~Department of Physics, Wabash College, 
Crawfordsville, Indiana 46556, USA. }}

Collectively interacting colloidal particles   
are often used as model systems  
to investigate various features of 
equilibrium and non-equilibrium phenomena \cite{1}.   
Due to their size scale, colloids provide the advantage that
microscopic information  
on the individual particle level 
can be directly accessed, something which is 
normally difficult or impossible in smaller scale systems 
such as nanoparticles, molecules, or atoms \cite{2,3}.  
Additionally, there are various methods such as optical techniques \cite{4,5} 
for
controlling colloidal ordering and manipulating 
individual colloids.
Examples of phenomena that have been studied with colloids include
two-dimensional melting transitions \cite{6,7}, 
solid-to-solid phase transitions \cite{8}, glassy dynamics \cite{3},
commensurate and incommensurate phases \cite{9,10,11,12}, 
depinning behaviors \cite{13,14,15,16}, self-assembly \cite{17,18}, and dynamic sorting \cite{19,20,21}.
It is also possible to use colloids to study plastic
deformation under shear in 
crystalline \cite{22} or amorphous \cite{23} colloidal assemblies. 
In crystalline materials, plastic deformations occur via
the motion of dislocations which often occur  
in bursts of activity or avalanches  \cite{24,25,26,27,28}. 
Certain studies that may be difficult to undertake in other systems
become feasible to perform with 
colloidal systems, 
such as observations of
changes in the particle configurations and dynamics during compression.  
Due to the relative softness of 
charge-stabilized colloidal assemblies, compressional studies 
can be 
performed in this system over a wide range of parameters and packing densities. 
In recent experiments, Varshney {\it et al.} \cite{29} 
demonstrated that it is possible to create a 
quasi-two-dimensional 
colloidal raft system confined between two barriers, and to then
compress or decompress the raft in order to dynamically change the
packing density. 
In these experiments the colloid-colloid interaction was
more complex than a simple repulsion, so that transitions 
from a loose-packed 
amorphous solid to a
much denser but still amorphous solid 
were observed; however, during the compression, various
plastic rearrangements occurred 
that produced hysteresis 
across the compression and decompression cycle \cite{29}.  

Several studies have addressed order-disorder transitions 
of repulsive colloids confined
between two parallel walls, 
where the transitions are associated with 
changes in the number of rows
of colloids that fit between the walls. 
Crystalline states arise for integer numbers of rows,
while disordered states occur when partial rows, or
a non-integer number of rows, 
are present \cite{30,31}. 
It should be possible to create
a two-dimensional (2D) system of repulsively interacting 
colloids that tend to form a triangular lattice, confine
the system by barriers or an anisotropic trapping potential 
that can be dynamically changed 
to compress the colloidal assembly in one direction, 
and observe how the configuration changes 
under compression as the colloids 
rearrange in order to minimize the repulsive colloid-colloid interactions. 
For example,
in numerical studies of confined charged particles 
in anisotropic traps, the particle configuration
changes as a function of the trap width \cite{32,33,34}.  
In experiments on  
ion trapping systems where the ions 
are confined in a quasi-one-dimensional potential with a
tunable strength, 
structural changes
in the ion configuration were observed as the assembly was compressed and
decompressed,
such as a transition from a single row of ions 
to a zig-zag or kink state \cite{35,36}.
It should also be possible to create dynamical
confining potentials or 
barriers for dusty plasma crystals \cite{37}.        

In this work we numerically study a 2D system of 
colloids with long-range Coulomb repulsive interactions
placed in a quasi-one-dimensional
confining potential such that as the depth of the potential is increased, 
the colloidal assembly is 
compressed in one direction.
Since the colloids can minimize their interaction energy by
adopting a triangular ordering with equal 
lattice spacing in all directions,
the colloids adjust their positions to make the 
local lattice structure as isotropic as possible during compression. 
Colloidal motion occurs both through
slow elastic distortions and via abrupt
avalanches in which plastic rearrangements   
occur. 
During the avalanche events, the average distance 
between nearest neighbor colloids increases.
Dislocations can be created or annihilated 
at the sample surface during the plastic events,
causing the colloids to move along shear bands.
For high compressions,
we find that avalanches are associated with reductions
of the number of rows
of colloids that fit in the confining potential.
Between the row reduction events, the colloids can adopt
partial crystalline or smectic order.
We also find that the colloid velocity distribution
during the avalanche events is non-Gaussian 
with a power-law tail, and that the exponents are consistent  
with those observed for velocity distributions in 2D dislocation 
systems \cite{32,38,39}. 
Under decompression, we observe a similar set of dynamics and 
find some hysteresis between the compression 
and decompression cycles; however, there are no large-scale hysteretic 
effects since the particle-particle interactions are purely repulsive.

\begin{figure}
\centering
\includegraphics[width=0.4\textwidth]{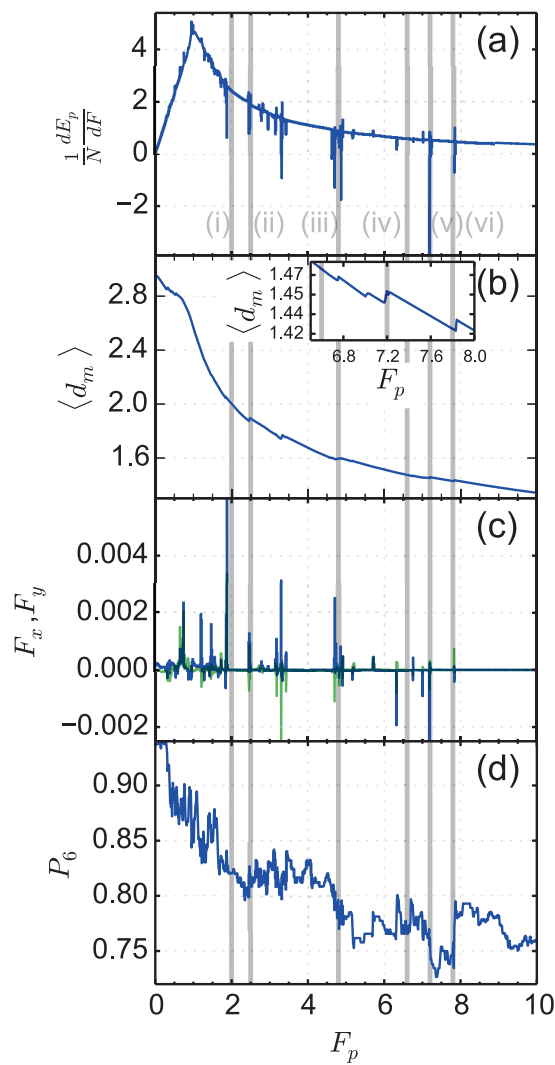}
\caption{
Under compression, achieved by increasing $F_p$, the particle motion occurs 
through both slow elastic distortions as well sudden plastic events or
avalanches that are associated with 
spikes or jumps in the following quantities: 
(a) $N^{-1}dE/dF_{p}$, the change in the particle configuration energy, 
vs $F_{p}$. 
(b) $\langle d_m\rangle$, the average distance between nearest-neighbor
particles,
vs $F_{p}$. 
Inset: a blowup of the main panel 
showing that plastic events are associated with increases in 
$\langle d_{m}\rangle$.  
(c) $F_x$ (blue) and $F_y$ (green), the 
average transient
forces on the particles in the $x$ and $y$ directions, vs $F_{p}$.  
(d) $P_6$, the fraction of six-fold coordinated particles, 
vs $F_{p}$.  Grey bands across all panels correspond to the intervals
of $F_{p}$ from which the particle trajectories in Fig.~\ref{fig:2} 
are taken.}  
\label{fig:1}
\end{figure}

\section{Simulation and System}   
We consider a 
two-dimensional assembly of 
colloidal particles interacting repulsively via  
a 
long-range Coulomb potential. 
We employ periodic boundary conditions in both the $x$
and $y$-directions for a system of size $L_x \times L_y$.
The sample contains $N=256$ colloids and
a single trough potential which produces a force 
${\bf F}^{P}_i = F_{p}\cos(2\pi x_i/L_x){\hat {\bf x}}$.
As $F_{p}$ is increased, 
the particles are forced closer together in the $x$-direction. 
The dynamics of a single colloid $i$ 
are obtained by integrating the following
overdamped equation of motion:
\begin{equation}  
\eta \frac{d {\bf R}_{i}}{dt} = {\bf F}_{\rm tot}^{i} =
-\sum_{i\ne j}^{N}{\bf \nabla}V(R_{ij}) +  {\bf F}^{P}_{i}  \ .
\end{equation} 
Here $\eta$ is the damping constant,  
${\bf R}_{i(j)}$ is the position of particle $i(j)$,
$R_{ij} = |{\bf R}_{i} - {\bf R}_{j}|$,
and the particle-particle interaction potential is 
$V(R_{ij}) =  q^2E_{0}/R_{ij}$,  
where $E_{0} = Z^{*2}/4\pi\epsilon\epsilon_{0}a_{0}$,  
$q$ is the dimensionless interaction strength,
$Z^{*}$ is the 
effective charge of the colloid,
and $\epsilon$ is the solvent dielectric constant. 
We treat the long-range image interactions using a
real-space Lekner summation method \cite{Lekner}.
Lengths are measured in units of $a_{0}$, time in units of 
$\tau = \eta/E_{0}$, and forces in units of $F_{0} = E_{0}/a_{0}$.
We increase $F_{p}$ in 
small increments of $0.001$ over the range $F_p=0$ to 10 in
the first part of the work, 
and in increments of $0.01$ from $F_p=0$ to 100 in the second part 
of the work.
During each increment we measure the 
average transient force response on 
each particle in the two directions 
$F_{x}=\sum_{i}^{N}\sum_{t=t_0}^{t_0+\Delta t}{\bf F}_{\rm tot}^i(t) \cdot {\bf \hat x}$ and 
$F_{y}=\sum_{i}^N\sum_{t=t_0}^{t_0+\Delta t}{\bf F}_{\rm tot}^i(t) \cdot {\bf \hat y}$, 
along with the changes in the particle-particle interaction energy
$E=\sum_{i}^{N}\sum_{j\neq i}^{N}V(R_{ij})$, 
the total energy of the system
$E_{\rm tot}=E+\sum_{i}^{N}F_p\sin(2\pi x_i/L_x)$,
local ordering $P_6=N^{-1}\sum_{i=1}^N\delta(z_{i}-6)$, where $z_i$ is the
coordination number of particle $i$ obtained from a Voronoi tesselation,
and $\langle d_m\rangle=N^{-1}\sum_{i}^Nd_m^i$, where
$d_m^i$ is the spacing between particle $i$ and its nearest neighbor
as identified from a Voronoi tesselation.  

\begin{figure}
\centering
\includegraphics[width=0.5\textwidth]{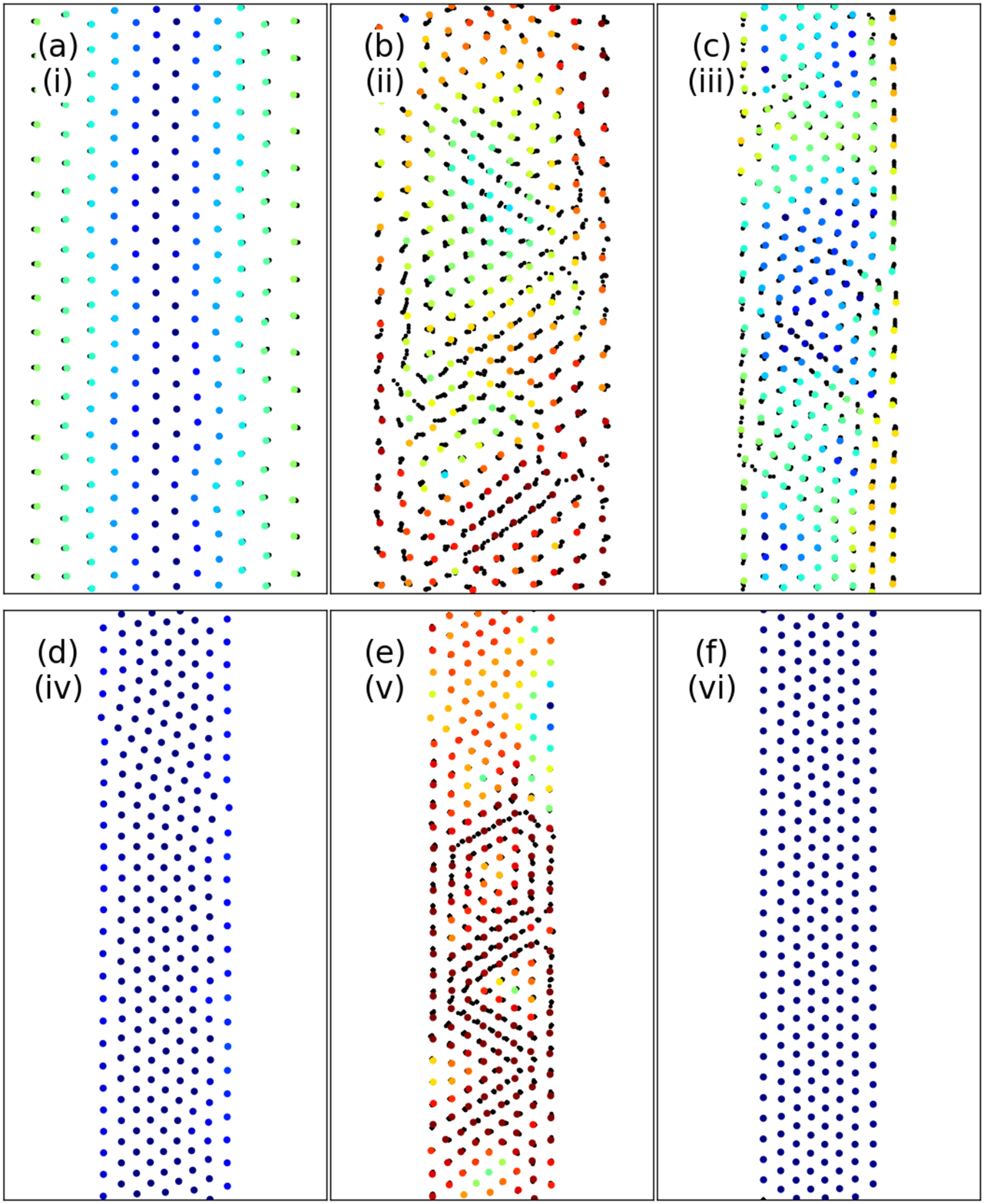}
\caption{
The particle positions (colored dots) in the 
potential as the particle assembly is compressed along the $x$ direction by 
increasing $F_p$.
Blue particles are stationary or slowly moving while red particles are
moving the most rapidly.
Small black dots indicate the particle trajectories
over a force interval of 
$\delta F_p=0.1$. 
(i) At $F_{p} = 2.0$, the motion is predominantly elastic. 
(ii) At $F_{p} = 2.5$, an avalanche occurs in the form
of shear bands. (c) 
At $F_{d}= 4.8$ the avalanche activity is more localized.
(d) At $F_{p} = 6.6$ only elastic  motion occurs.
(e) At $F_{p}= 7.2$ there is a large avalanche. 
(f) At $F_{p} = 8.0$ the motion is predominantly elastic.  
}
\label{fig:2}
\end{figure}
   
\section{Compression} 
In Fig.~\ref{fig:1}(a) 
we plot the change in the energy of the particle configuration 
$N^{-1}dE/dF_{p}$ versus $F_{p}$. 
Figure~\ref{fig:1}(b) shows the corresponding $\langle d_m\rangle$, the
average spacing between nearest-neighbor particles, 
Fig.~\ref{fig:1}(c) shows the
cumulative transient forces $F_{x}$ and $F_{y}$, 
and Fig.~\ref{fig:1}(d) 
shows the fraction of six-fold coordinated particles $P_{6}$. 
In Fig.~\ref{fig:2} we plot
the particle positions and trajectories over
intervals of $\delta F_{p} = 0.1$ which correspond
to the gray shaded bands in Fig.~\ref{fig:1}.
We find two distinct types of behaviors.  Elastic distortions occur in
the smoothly changing portions of $dE/dF_{p}$ and $\langle d_{m}\rangle$, in
the same regions where $F_{x}$ and $F_{y}$ are close to zero. 
Sudden plastic events or avalanches are associated 
with peaks in $dE/dF_{p}$, jumps in 
$\langle d_{m}\rangle$, and spikes in $F_{x}$ and $F_y$. 
The particle configuration energy $E$ increases with 
increasing $F_{p}$ as the repulsively interacting
colloids are forced together, 
which accounts for the overall positive value of  
$dE/dF_{p}$. 
The cusp in $dE/dF_{p}$ near $F_{p} = 1.0$ appears when
$F_{p}$ becomes large enough that the particles can no longer spread to
cover the entire sample but instead segregate into the bottom of the
potential, with a particle-free region appearing at the potential maximum that
becomes wider with increasing $F_p$.
\begin{figure*}
\centering
\includegraphics[width=0.65\textwidth]{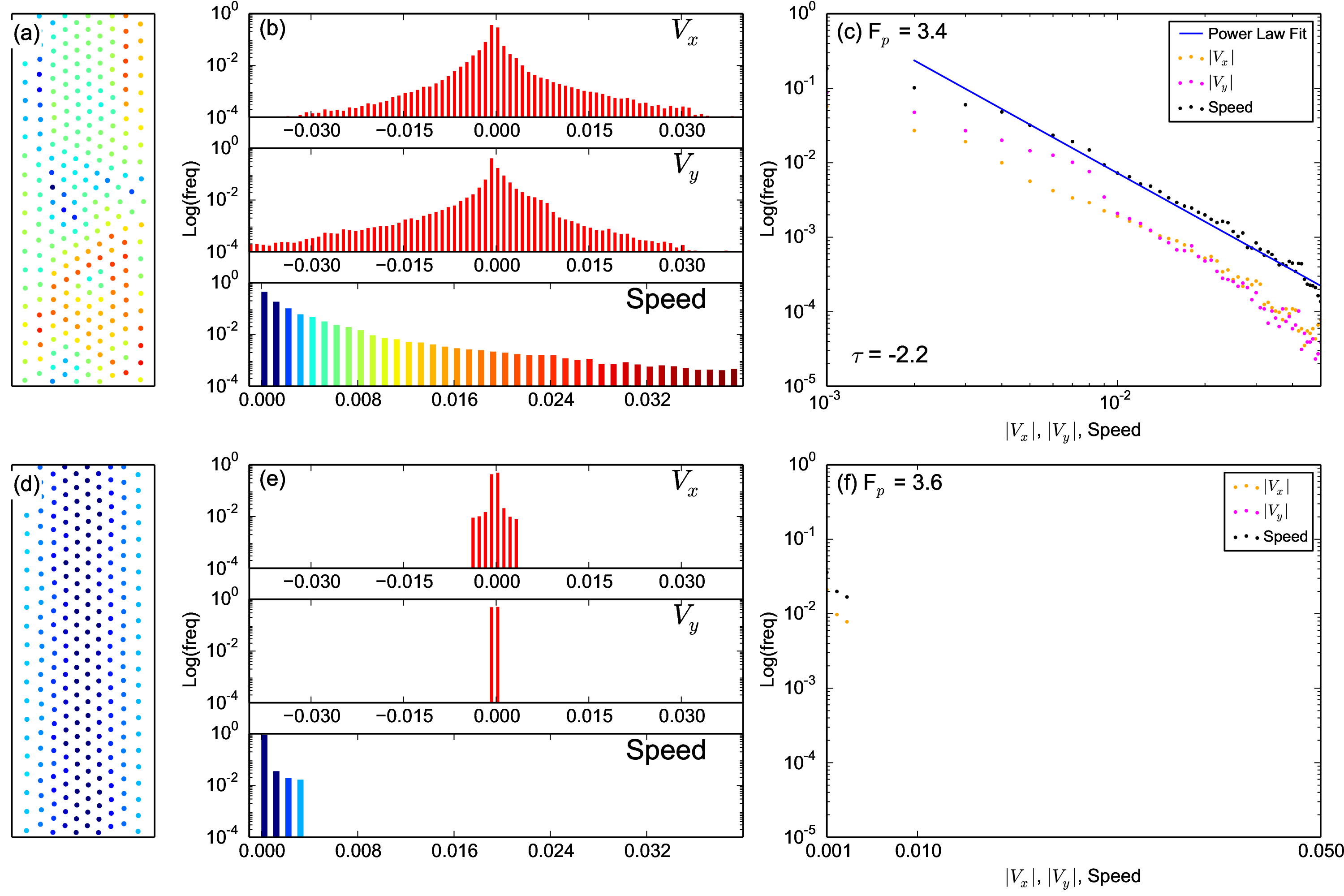}
\caption{(a,d) Particle positions colored by the amount the particle
moved during an increment of $F_p$, with blue particles nearly stationary and
red particles moving the largest amount.
(b,e) 
Velocity distributions $P(V_x)$, $P(V_y)$, and $P(|V|)$, from top to
bottom.  The same color scale shown for $P(|V|)$ is used in panels (a)
and (d).
(c,f) 
$P(|V_x|)$, $P(|V_y|)$, and $P(|V|)$ plotted on a log-log scale.  The solid
line in panel (c) is a power-law fit $P(|V|) \propto |V|^\tau$
with an exponent of $\tau=-2.2$.
Panels (a,b,c) are for a force interval of $3.4 < F_p < 3.5$ where
an avalanche occurs, and
panels (d,e,f) are for a force interval of $3.6 < F_p < 3.7$ where the
behavior is elastic and the distributions are much sharper.
}
\label{fig:3}
\end{figure*}

The spikes in $dE/dF_{p}$ generally have negative values,
indicating that they are associated with
drops in the configuration energy.
The plastic events are 
also associated with an increase in the average spacing between
nearest-neighbor particles, as indicated in Fig.~\ref{fig:1}(a,b)
where negative spikes in $dE/dF_{p}$ correlate with positive jumps 
in $\langle d_{m}\rangle$. 
The inset in Fig.~\ref{fig:1}(b) shows a blowup of the 
plastic events near $F_{p} = 7.2$ to better highlight 
the discrete nature of the jumps in $\langle d_m\rangle$.
Since the colloid-colloid interactions are repulsive, 
the configurational energy increases under compression as the particles
elastically approach each other along the compressed direction,
while during the plastic avalanches, 
the closest spacing between particles increases so that
the configurational energy is reduced.
In Fig.~\ref{fig:1}(c) the peaks in $F_x$ coincide with
peaks in $F_y$, indicating that the particle 
motion is occurring almost equally in both
the $x$ and $y$ directions during an avalanche. 
Fig.~\ref{fig:1}(d) shows that initially
$P_{6}\approx 0.95$
at $F_{p} = 0.0$ where the system forms 
a slightly distorted ordered triangular lattice. 
As $F_{p}$ increases, there 
are numerous very small peaks in $P_6$ caused by
the Voronoi construction having difficulties with the sample edges; 
however, there 
are several real larger scale jumps in $P_{6}$ that are
correlated with the
avalanche events.  
The increases and decreases in $P_6$ occur when dislocations, which have 
fivefold or sevenfold rather than sixfold coordination, are created or 
annihilated
during an avalanche.

In the avalanche illustrations of Fig.~\ref{fig:2},  
red particles 
move significantly during the compression increment
while blue particles experience little to no motion.
Figure~\ref{fig:2}(a) shows the state near 
$F_{p} = 2.0$ for an interval in which there
are no spikes or jumps in the quantities 
plotted in Fig.~\ref{fig:1}, indicating that the 
particles are undergoing elastic motion. 
Here the particles are arranged in twelve vertical rows.
The avalanche near $F_p=2.5$ is shown in 
Fig.~\ref{fig:2}(b)
where the particles transition from twelve to eleven vertical rows.
Here, motion occurs in localized bands, indicative of 
a shear banding effect. 
These bands generally form zig-zag patterns,
and there are also regions undergoing local rotation.      
Figure~\ref{fig:2}(c) illustrates another avalanche 
interval near $F_{p} = 4.8$, where
similar shear banding motion appears.
In this case, 
there is considerable motion of the particles
on the outer edges of the sample 
and the particles transition from ten to nine vertical rows.
Figure~\ref{fig:2}(d) shows another elastic region 
near $F_{p} = 6.6$ where there is little motion. 
Near $F_{p} = 7.2$, a large plastic event occurs as illustrated
by the bands of motion in Fig.~\ref{fig:2}(e) 
as the system transitions from nine to eight rows.
For $7.2 < F_p < 10$, 
the system
behaves elastically, as shown in Fig.~\ref{fig:2}(f) 
for $F_{p} = 8.0$ where little motion occurs. 
Further compression of the system beyond $F_p=10$ is described in 
Section~\ref{sec:4}.    

\subsection{Velocity Distributions}

We next examine histograms of the particle velocities 
at different values of $F_{p}$. 
Figure~\ref{fig:3}(a) shows the particle positions
for $3.4 < F_{p} < 3.5$, where the color denotes the average speed
$|V|=\sqrt{V_x^2+V_y^2}$ of each particle, 
highlighting the heterogeneous displacements of particles in bands.  
In Fig.~\ref{fig:3}(b) we plot $P(V_x)$, $P(V_y)$, and $P(|V|)$ for
the same force interval.
The distributions are obtained by averaging the velocities of each particle
over ten subintervals of $\delta F_p=0.01$. 
The plastic motion that occurs during the avalanche event produces
non-Gaussian velocity distributions, as shown more clearly
in Fig.~\ref{fig:3}(c) where we plot $P(|V_x|)$, $P(|V_y|)$, and $P(|V|)$
on a log-log scale.
The solid line is a power law fit of
$P(|V|) \propto |V|^\tau$, where $\tau = -2.2$.    
In Fig.~\ref{fig:3}(d,e,f) we show the same 
quantities for $3.6 < F_{p} < 3.7$ which corresponds to
an interval in which the response is elastic, as indicated 
in Fig.~\ref{fig:3}(d).
The width of the velocity distributions in Fig.~\ref{fig:3}(e)
is significantly smaller than for the plastic flow event illustrated
in Fig.~\ref{fig:3}(b).
Due to the sharpness of the distributions, 
the plot of $P(|V|)$ on a log-log scale in Fig.~\ref{fig:3}(f) 
cannot be fit to a power law.   
In 
Fig.~\ref{fig:3}(g,h,i)
we show the same quantities for 
$7.1 < F_{p} < 7.2$,  corresponding to an interval 
containing an avalanche as highlighted in Fig.~\ref{fig:1}.
Here the motion follows a banding pattern.  
The velocity distributions have strongly non-Gaussian
features, as shown in 
Fig.~\ref{fig:3}(g,h).
A fit of $P(|V|) \propto |V|^\tau$
in 
Fig.~\ref{fig:3}(i)
gives
$\tau = -1.9$. 
For $7.3 < F_{p} < 7.8$ there are few plastic rearrangements and the velocity
distributions are very similar to those
shown in 
Fig.~\ref{fig:3}(e,f)
for the $F_{p}= 7.0$ case. 
In general, we find that during the avalanches, 
the velocity distributions of the particles
have a non-Gaussian form with a power law tail 
that can be fit with $1.9 < \tau < 2.6$, while in the elastic regimes, 
the width of the velocity distributions is strongly reduced. 

\begin{figure}
\centering
\includegraphics[width=0.5\textwidth]{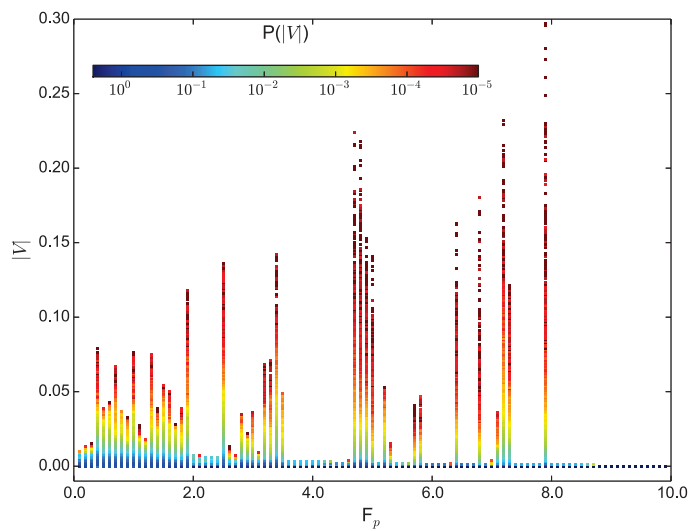}
\caption{ A heightfield plot of $P(|V|)$ vs $F_p$, where as indicated
by the scale bar blue values of $|V|$ occur with high frequency while
red values of $|V|$ occur with low frequency.  The total height of
each vertical column indicates the maximum range of $|V|$ for that value
of $F_p$.
The maximum value of this range gradually increases with increasing $F_p$.
}
\label{fig:5}
\end{figure}

Previous studies of 
sheared crystalline materials 
showed that avalanches are associated with the motion of dislocations 
\cite{24,25,26,27,28,32,38,39}, and that various quantities such as
the dislocation velocities are power-law distributed in the avalanche
regime.   
We consider the particle velocities instead of the dislocation velocities
in our system;
however, 
due to the partial crystalline order of our sample,
avalanches are generally associated with the motion of dislocations. 
In the dislocation dynamics studies, the high velocity tails of the
velocity distributions can be fit with power law exponents of
$\tau = -2.5$ \cite{32} and  $-3.0$ \cite{38}. 
Recent computational and theoretical work showed that in 2D,
a single dislocation has a
power-law distributed velocity
with $\tau = -2$, while when collective effects are important,
larger exponents of $\tau = -2.4$ appear \cite{39}. 
In our case, as dislocations move through the system, individual particles
temporarily translate along with the dislocations, so that over
sufficiently short time windows
the overall form of the velocity
distributions of the particles and the dislocations should be
similar.
It is difficult to extract the exact exponents for our 
system due to the finite width
of the sample and the fact that some  avalanches are 
associated with only one moving dislocation while others contain multiple 
moving and interacting dislocations. 
The width of the distribution is also dependent on the packing fraction.
As the particle density increases, the maximum possible particle velocity 
also increases.  
This is shown in Fig.~\ref{fig:5} where we plot a color height map of $P(|V|)$
versus $F_p$ for intervals of $\delta F_p=0.1$, highlighting the maximum
extent of the high-velocity tails of the distributions.
As $F_p$ increases, the particle packing density increases and the
maximum achievable particle velocity also increases significantly.

\begin{figure}
\centering
\includegraphics[width=0.5\textwidth]{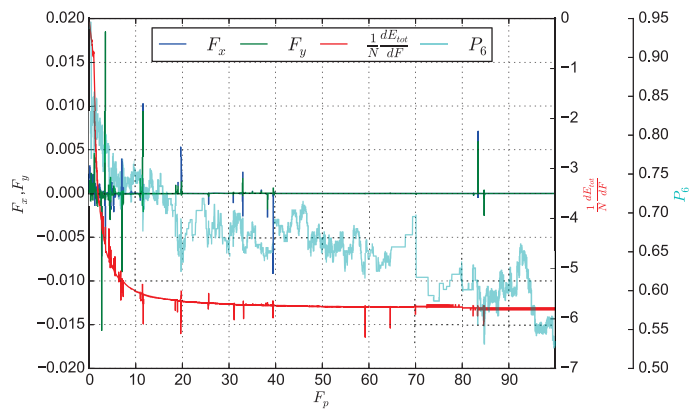}
\caption{ $F_{x}$, $F_{y}$ $N^{-1}dE/dF_p$, and $P_{6}$ for a system 
subjected to a maximum compression of
$F_{p} = 100$. The density of avalanches is much lower for $F_p>10$ than
in the $F_p<10$ regime already discussed,
and the avalanches at higher $F_p$ are associated with a partial or 
complete reduction in the number of rows of
particles in the system.
}
\label{fig:6}
\end{figure}

\begin{figure}
\centering
\includegraphics[width=0.5\textwidth]{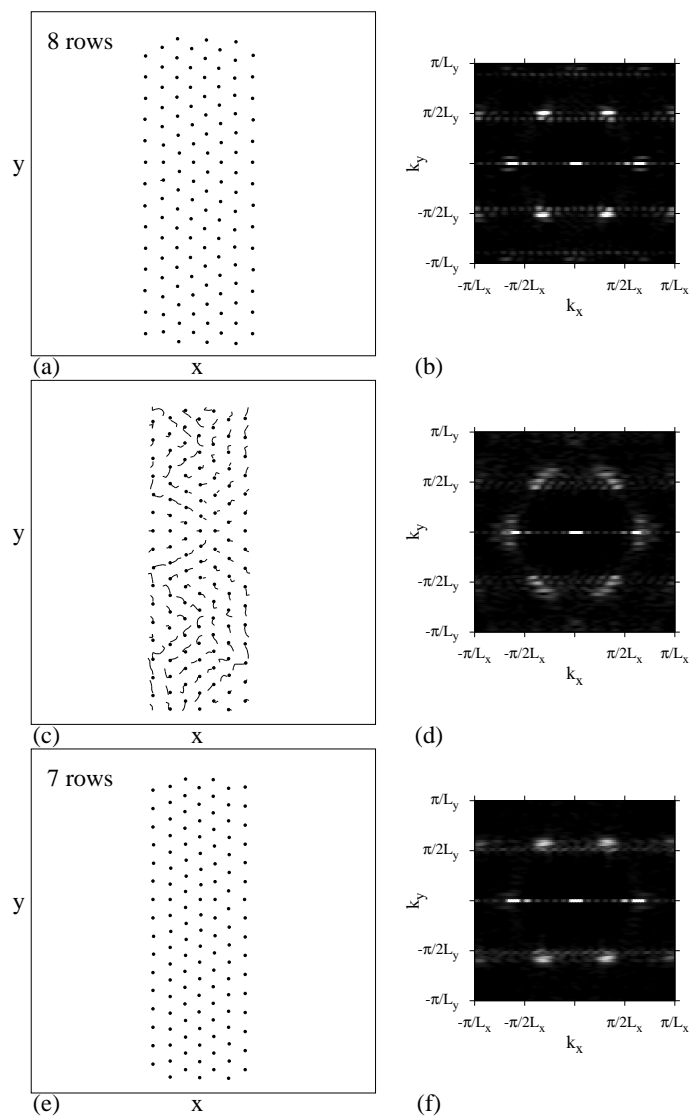}
\caption{ 
(a) Real space image of particle locations 
in a portion of the sample at $F_p=10$
showing that there are eight rows of particles. 
(b) The corresponding structure factor $S({\bf k})$ at $F_{p} = 10$
indicates strong triangular ordering.
(c) Particle positions and trajectories 
in a portion of the sample during an 
avalanche event at $F_{p} = 11.6.$ (c) $S({\bf k})$
during the avalanche shows a disordered structure.
(e) Real space image of the particles 
in a portion of the sample at $F_{p} = 11.8$ 
where there are now seven rows of particles. (f) The corresponding
$S({\bf k})$ shows triangular ordering.  
}
\label{fig:7}
\end{figure}

\begin{figure}
\centering
\includegraphics[width=0.5\textwidth]{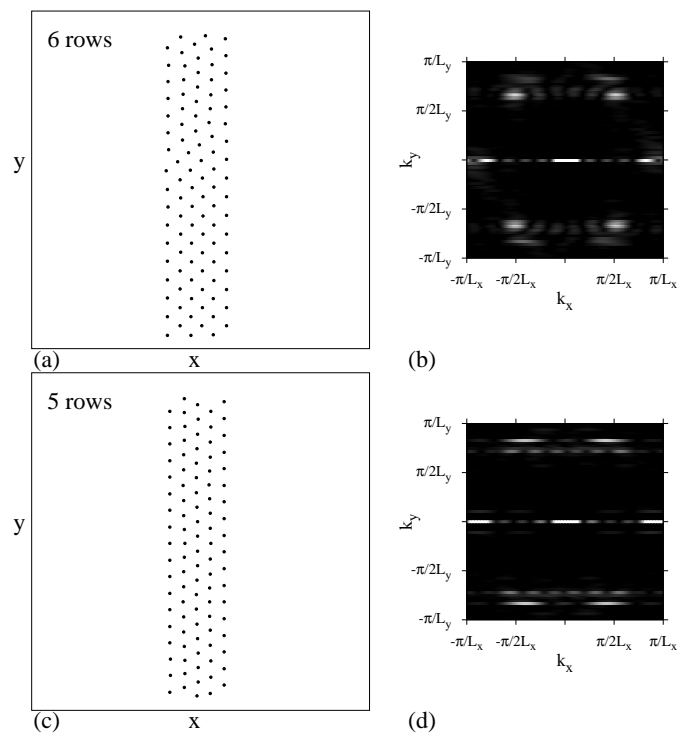}
\caption{(a) Real space image of particle locations in a portion of
the sample at $F_p=39$, where there are six rows of particles.
(b) The corresponding $S({\bf k})$.
(c) Real space image of particle locations in a portion of the sample
after an avalanche at $F_p=40$.  There are now five rows of particles.
(d) The 
corresponding $S({\bf k})$ is more
smeared, indicating a smectic-like structure.  
}
\label{fig:8}
\end{figure}

\begin{figure}
\centering
\includegraphics[width=0.5\textwidth]{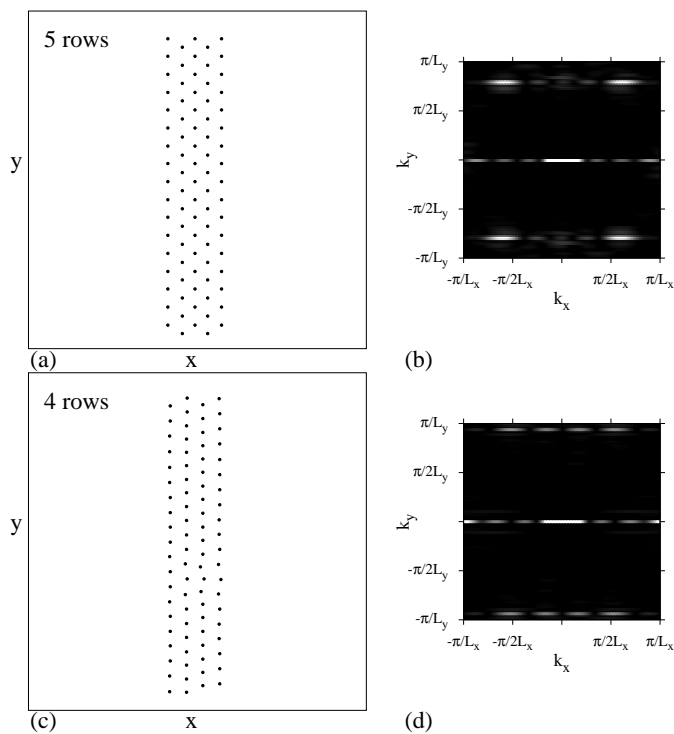}
\caption{(a) Real space image of particle locations in a portion of
the sample at $F_p=80$, where there are five rows of particles.  (b) The
corresponding $S({\bf k})$.  
(c) Real space image of particle locations in a portion of the sample after
an avalanche at $F_p=87$.  There are now five rows of particles.
(d) The corresponding $S({\bf k})$ has a smectic structure.
}
\label{fig:9}
\end{figure}

\section{Row Reduction Transitions\label{sec:4}} 

In Fig.~\ref{fig:6} we plot simultaneously
$F_{x}$, $F_{y}$, the derivative of the total energy $dE_{\rm tot}/dF_{p}$, 
and $P_{6}$ versus $F_{p}$ as the sample is compressed all the way to
$F_{p}=100$.
The density of avalanche-induced peaks is much smaller for $F_p>10$ than
for $F_p<10$, while
the avalanche jumps at higher $F_p$ values clearly coincide
with sharp changes in $P_{6}$. 
In general, plastic avalanche events that occur for $F_p>10$
are associated with a partial or complete reduction of the 
number of rows of particles
that fit across the potential well.
In Fig.~\ref{fig:7}(a) we show the particle positions in a portion of
the sample at $F_p=10$, with the corresponding structure
factor $S({\bf k})=N^{-1}|\sum_i^Ne^{-i{\bf k}\cdot{\bf r}_i}|^2$
plotted in 
Fig.~\ref{fig:7}(b). 
Here, eight rows of particles 
fit across the potential well and
the system has 
significant triangular ordering as indicated by the
six-fold peaks in $S({\bf k})$. 
Figure~\ref{fig:7}(e,f) shows the real space particle positions and
$S({\bf k})$ at
$F_{p} = 11.8$, where there 
are only seven rows of particles that still have
mostly triangular ordering. 
Figure \ref{fig:6} shows that between $F_{p} = 10$ and 
$F_p=11.8$ a plastic avalanche event occurs,
as indicated by the spikes in $F_{x}$, $F_y$, and 
$dE_{\rm tot}/dF_{p}$. 
We illustrate the correlated bands of particle motion that occur
during the
$F_p=11.56$ event in 
Fig.~\ref{fig:7}(c). 
Figure~\ref{fig:7}(d) shows 
that during the avalanche, the
sixfold peaks in $S({\bf k})$ 
are heavily smeared, 
indicating the disordering of the structure as dislocations move
through the system. 
We observe similar $S({\bf k})$ signatures for avalanches at higher $F_{p}$. 
In some cases, the number of rows is not reduced uniformly
through the entire system in a single avalanche; instead,
a portion of the system collapses to the smaller number of rows in
one avalanche event, followed by a second avalanche event at
a slightly higher value of $F_p$ that completes the row reduction process.

\begin{figure}
\centering
\includegraphics[width=0.4\textwidth]{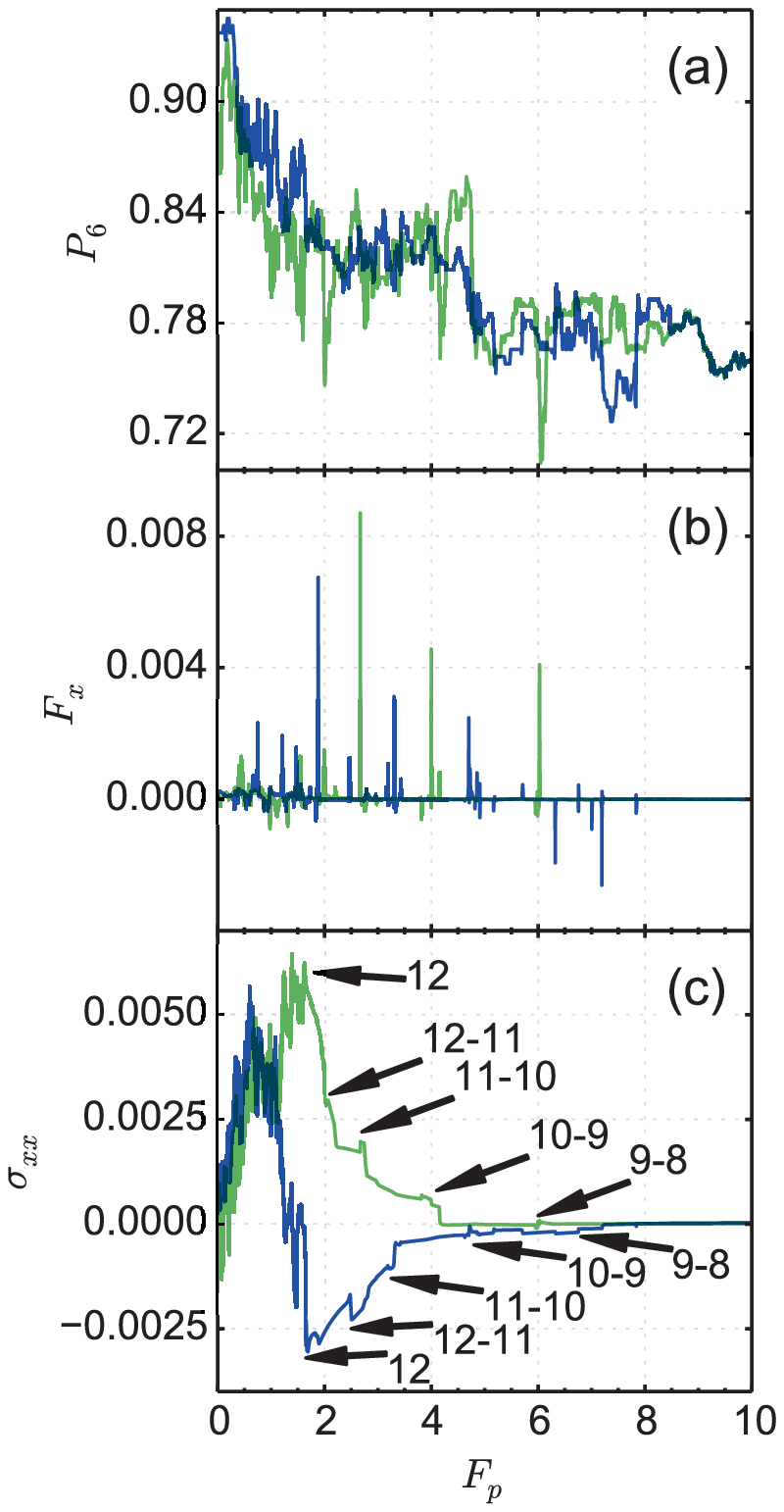}
\caption{ 
Hysteretic response measurements.  Blue curves are obtained during
the initial increasing sweep of $F_p$, and green curves are from
the decreasing sweep in which $F_p$ is reduced back to zero.
Grey bands are at the same locations as in Fig.~\ref{fig:1}, and
indicate the intervals during the increasing sweep of $F_p$ at
which the particle trajectories in Fig.~\ref{fig:2} were obtained.
(a) $P_6$ vs $F_p$.  The blue curve for increasing $F_p$ also
appears in Fig.~\ref{fig:1}(d).
(b) $F_x$ vs $F_p$.  The blue curve for increasing $F_p$ also
appears in Fig.~\ref{fig:1}(c).
(c) $\sigma_{xx}$ vs $F_p$.
}
\label{fig:10}
\end{figure}

\begin{figure*}
\centering
\includegraphics[width=0.65\textwidth]{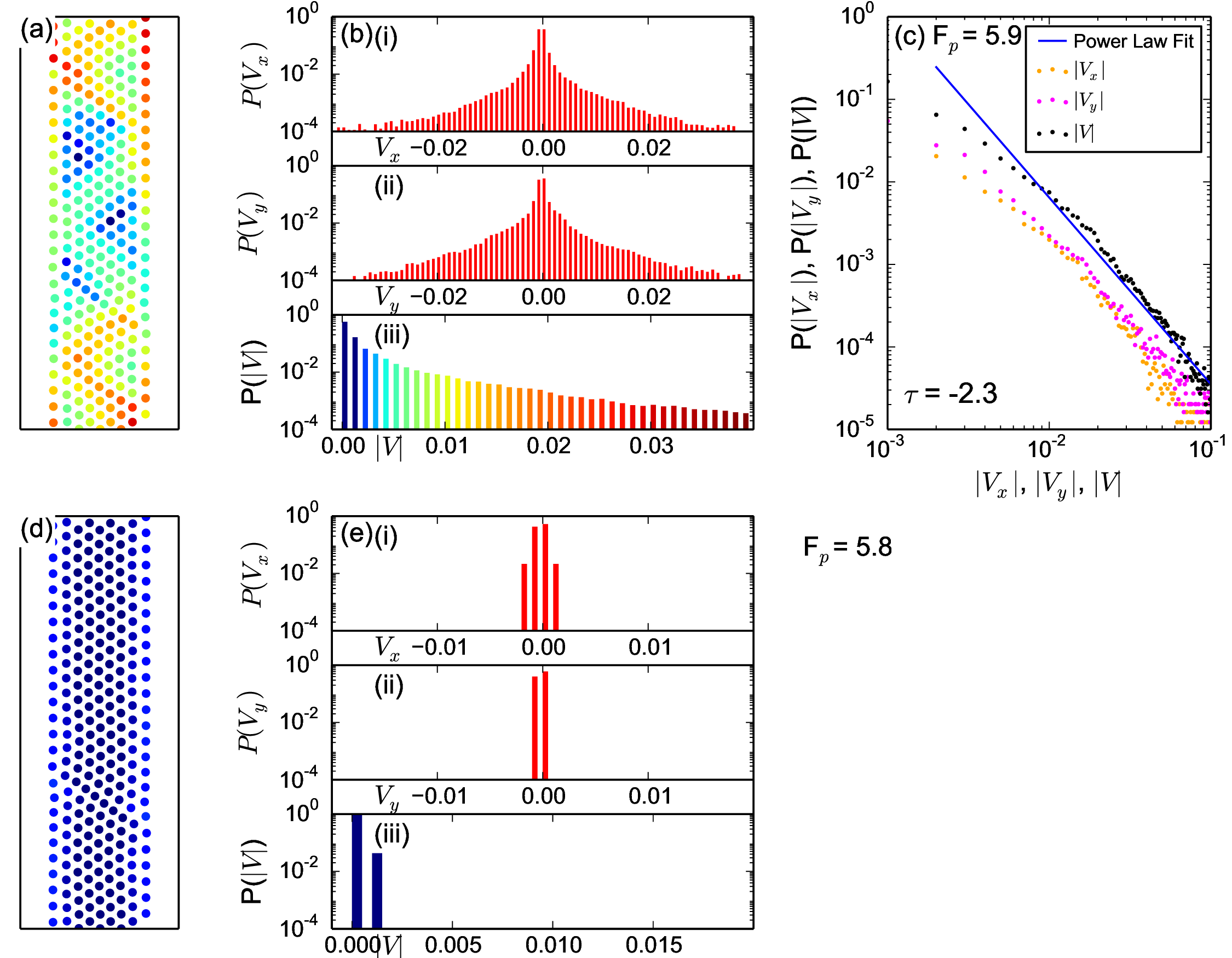}
\caption{
(a,d) Particle positions colored by the amount the particle moved during
an increment of $F_p$, with blue particles nearly stationary and red
particles moving the largest amount.  (b,e) Velocity
distributions $P(V_x)$, $P(V_y)$, and $P(|V|)$, from top to bottom.
The same color scale shown for $P(|V|)$ is used in panels (a) and (d).
(c) $P(|V_x|)$, $P(|V_y|)$, and $P(|V|)$ plotted on a log-log scale.
Solid line is a power-law fit of $P(|V|) \propto |V|^{-\tau}$, where
$\tau=-2.0$.
Panels (a,b,c) are for a force interval of $5.9 > F_p > 5.8$ during
decompression, where a large plastic event occurs.
Panels (d,e) are for a force interval of $5.8 > F_p > 5.7$ during
decompression, where the system behaves elastically.
}
\label{fig:11}
\end{figure*}

In Fig.~\ref{fig:8}(a,b) we show the real space and 
$S({\bf k})$ plots for a sample at $F_{p} = 39$ 
where there are six rows of particles. 
Here
there are patches in the system where the triangular ordering has been
distorted into a rhomboidal ordering, as seen by the nonuniform spacing
between the peaks in $S({\bf k})$.
Figure~\ref{fig:6} shows that an avalanche occurs near 
$F_{p} = 40$. 
After the avalanche, at $F_p=44$,
the real space 
plot in 
Fig.~\ref{fig:8}(c) 
has only five rows of particles, 
indicating that the avalanche
was associated with a row reduction. 
The structure in Fig.~\ref{fig:8}(c) has a smectic characteristic 
as indicated by the one-dimensional smearing of the peaks 
along the $k_{x}$ direction in the $S({\bf k})$ plot in Fig.~\ref{fig:8}(d).
For $42 < F_{p} < 80$, the system retains five rows of 
particles; however, there are still avalanche 
events associated 
with the system shifting between different partially crystalline structures
that still contain five rows.
For example, 
Fig.~\ref{fig:9}(a) shows a rhomboidal ordering of the 
five rows of particles at
$F_p=80$, and Fig.~\ref{fig:9}(b) shows the corresponding $S({\bf k})$.
In the range $80 < F_{p} < 87$, several avalanche events 
occur that are associated with 
a transition from five to four rows of particles. 
In Fig.~\ref{fig:9}(c,d) we show the particle positions
and $S({\bf k})$ at $F_{p} = 87$, where there are four rows of particles 
in a smectic arrangement.

\section{Decompression and Hysteresis}

To explore hysteretic effects,
once the system is fully compressed to 
$F_{p} = 10$ we gradually decompress the sample by decrementing $F_p$ back
to zero, using the same sweep rate for $F_p$ 
as during the original compression.
We find that, just as during compression, under decompression
the system undergoes
combinations of slow elastic distortions 
interspersed with sudden avalanche rearrangement events. 
In Fig.~\ref{fig:10}(a,b) we plot $P_6$ and  $F_x$ 
versus $F_p$ for the compression and decompression cycles in a sample where
the maximum compression value is $F_p=10$.
Figure~\ref{fig:10}(a) shows that the $P_{6}$ curves 
for compression and decompression overlap completely above
$F_p \sim 8.5$, while in Fig.~\ref{fig:10}(b) there are no
avalanche events in this same force interval, indicating that
the system is behaving reversibly.
Avalanche events are completely suppressed during the decompression until
$F_p$ has decreased to $F_p=6.0$, where the first large peak in $F_x$ appears
in Fig.~\ref{fig:10}(b) accompanied by a large dip in $P_6$ in 
Fig.~\ref{fig:10}(a).
During this event, a number of dislocations enter the system and the
average spacing between the colloids increases. 
In general we observe fewer plastic events 
during the decompression cycle than during compression; however, the
plastic events that do occur during decompression tend to be larger. 
Two large
plastic decompression events 
appear at
$F_{p} = 4.0$ and $F_p=2.6$, marked by spikes in $F_x$ and substantial
drops in $P_6$.
We find 
no systematic trend in the hysteresis 
such as $P_6$ or $F_x$ always being higher for one direction of the
compression cycle.
Instead, 
the plastic avalanches
occur at different values of $F_{p}$ for each driving direction,
indicating the irreversibility associated with the plasticity in this
system. 
The lack of systematic hysteresis is different from the behavior 
observed for compression-decompression cycles in 
colloidal raft experiments \cite{29}. 
In those experiments, the particle-particle interactions had both
repulsive and attractive components,
so that during the initial compression the particles could
be captured by the attractive portion of the interparticle potential.
In contrast, in our system all of the particle-particle interactions are
smoothly repulsive.

\begin{figure}
\centering
\includegraphics[width=0.5\textwidth]{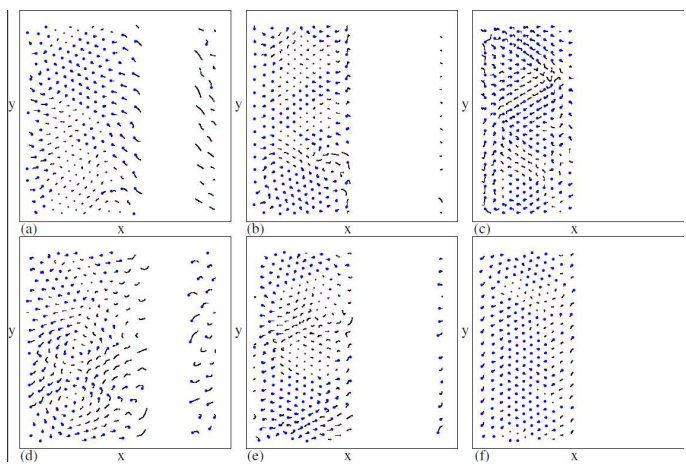}
\caption{
Images illustrating the hysteretic behavior
of the sample shown in Fig.~\ref{fig:10}(c) over the range
$F_p=1.1$ to $F_p=1.8$
where the system transitions from diagonal to vertical rows of particles.
Particle positions are colored by the amount the particle moved during
an increment of $F_p$, with blue particles nearly stationary and red
particles moving the largest amount. 
Small black dots indicate the particle trajectories
over a force interval of $\delta F_p = 0.1$.  
In (a-c) the system is being compressed while in (d-f) the system 
is being decompressed.
(a,d) At $F_p = 1.1$ the rows of particles are not oriented along the $y$
axis.
(b,e) At $F_p = 1.6$ the rows in the compressing system (b) are vertical
while those in the decompressing system (e) are diagonal.
(c,f) At $F_p = 1.8$ the rows are vertical during both compression
and decompression. 
}
\label{fig:13}
\end{figure}

Figure~\ref{fig:10}(c) shows $\sigma_{xx}$, 
an element of the stress tensor: 
\begin{equation}
\sigma_{xx} = \frac{1}{S_x^2} \sum_{i}^{N}\sum_{j<i}{ (F_{ij})_x (\vec{r}_{ij})_x}
\end{equation} 
plotted versus $F_p$ for both compression and decompression. 
Here $\vec{F_{ij}}$ 
is the interparticle force between particles $i$ and $j$
and $\vec{r}_{ij}$ is their relative separation.    
There is no significant difference between the compression and decompression
values of $\sigma_{xx}$ over the range 
$F_p \geq 8.0$, 
but we find significant hysteresis for $F_p<8.0$.
For $F_p<1.8$, the $\sigma_{xx}$ signal is noisy under both
compression and decompression.  At very low $F_p$, the
particles are spread out across the entire sample, with no gap. 
At $F_p=1.5$, a particle-free 
gap opens at the location of the potential maximum, and
the particles form rows that are rotated by a finite angle from
the $y$ direction, as shown in Fig.~\ref{fig:13}(a,f).
This diagonal ordering reduces the energy of the system by 
minimizing the compression of the lattice, 
but when $F_p\geq 1.8$ this small energy advantage is lost and
the particles transition as illustrated in Fig.~\ref{fig:13} into 
a vertical row structure 
of the type shown in Fig.~\ref{fig:13}(c) that is aligned with the
direction of the confining potential.
For $1.8 < F_p < 8.0$, $\sigma_{xx}$ remains hysteretic but is
no longer noisy; instead, 
the curve is smooth with distinct jumps, 
indicating the hysteresis in the value of $F_p$ at which the
number of rows in the sample is reduced during compression or increased
during expansion.
The row transitions are marked by labeled arrows in Fig.~\ref{fig:10}(c),
where we observe an average offset of
$\Delta F_p \sim 0.7$ between the reduction in the number of rows from
$n$ to $n-1$ upon compression and the increase in the number of rows
from $n-1$ to $n$ upon decompression, with the transition during
compression falling at a higher value of $F_p$.
For example, the 11-10 transition during compression occurs at
$F_p = 3.2$ while the 10-11 transition during decompression occurs
at $F_p = 2.7$.

During decompression,
the particle velocity distributions in the elastic regimes and during
avalanches have
the same features described above for compression.
Figure~\ref{fig:11}(a,b,c) shows the particle configuration, 
velocity distributions, and log-scale plot of 
$P(|V_x|)$, $P(|V_y|)$, and $P(|V|)$ during the 
decompression cycle over the range
$5.9 > F_p > 5.8$. 
The velocity distributions are clearly non-Gaussian, 
and the power-law fit of $P(|V|)$ in Fig.~\ref{fig:11}(c)
has an exponent of $\tau = -2.4$.
Figure~\ref{fig:11}(d,e,f) shows the same 
quantities over the range
$5.8 > F_p > 5.7$ where elastic motion occurs.
Here the velocity distributions
are much sharper.

\section{Summary}

We have numerically investigated a monodisperse assembly of repulsively 
interacting Yukawa colloids
undergoing dynamical compression in a trough potential. During the compression
the motion is predominately elastic as
the particles gradually move closer together. 
This elastic motion is interspersed 
with occasional sudden plastic rearrangements or avalanches 
in which
the colloids shift position in order to increase the average 
spacing between neighboring particles.
The avalanches take 
the form of local shear banding 
events in which dislocations can be created or annihilated. 
During the plastic events, the colloidal velocity
distributions are non-Gaussian with power law
tails that have exponents ranging from $\tau = -1.9$ to $\tau=-2.5.$ 
This is consistent with the velocity distributions found for moving 
dislocations during 
avalanche events. 
For larger compressions, the avalanches are generally associated with 
row reduction events where the number of rows of colloids 
that fit inside the trough is partially or completely reduced
by one.  After these avalanche events, the colloids can exhibit
partial crystalline or smectic type structures. 
During decompression, we observe similar avalanche behaviors; 
however, the avalanche events do not occur
at the same values of substrate strength, indicating the occurrence
of irreversible behavior.
This system could be experimentally realized 
by confining colloids or other charged
particles such as dusty plasmas in anisotropic traps 
where the barriers or the trap width can be changed dynamically.

\section{Acknowledgements}
This work was carried out under the auspices of the 
NNSA of the 
U.S. DoE
at 
LANL
under Contract No.
DE-AC52-06NA25396.


%\balance

\footnotesize{
 
}
\end{document}